# Controlled transparency of many-mode waveguides with rough surface

**F. M. Izrailev[1] and N. M. Makarov[2]**

[1] Instituto de Fisica, Universidad Autonoma de Puebla, Apartado Postal J-48, Puebla, Pue. 72570, Mexico
[2] Instituto de Ciencias, Universidad Autonoma de Puebla, Priv. 17 Norte No 3417, Puebla, Pue. 72050, Mexico

In a unified approach we consider transport properties of 1D and quasi-1D waveguides with rough surfaces. Main attention is paid to the possibility of perfect transmission of waves due to specific long-range correlations in the surface profiles. First, we show how to construct random profiles that lead to a complete transparency of waveguides with one open channel. Then, we present analytical results for many-mode waveguides. It was revealed that by a proper choice of correlations in surface profiles the transmission through such quasi-1D waveguides is described by a coset of non-interacting 1D channels with a perfect transmission along each one. The number of these conducting modes is governed by the control parameter, and can be equal to the total number of channels. Therefore, the waveguides can be completely transparent in some region of frequency of incoming waves. This unexpected phenomenon is discussed in connection with the violation of the single-parameter scaling for surface scattering.

**1  1D waveguiding systems**  During last few years much attention has been paid to studying transport properties of one-dimensional (1D) systems with correlated disorder (see, e.g., Ref. [1] and references therein). The interest is due to intriguing results that may have potential applications both to electromagnetic and electron/optic devices. In particular, it was shown [2] that any desired combination of transparent and nontransparent frequency windows can be achieved by a proper construction of random potentials with specific long-range correlations. Experimental realization [3] of such potentials with delta-like scatters has confirmed the theoretical predictions for single-mode waveguides.

The subject of wave propagation through surface-corrugated waveguides is important for application in optics fibers, remote sensing, shallow water waves, etc., as well as in view of electronic transport in mesoscopic conducting structures. For this reason, we have recently analyzed anomalous properties of surface-governed transmission in waveguides with one open channel [4]. As a result, a theoretical method was proposed to construct disordered rough surfaces with specific long-range correlations along the waveguide. It was shown analytically and by direct numerical simulations as well, that single-mode waveguides with desired selective transparency can be fabricated with this method.

The idea of the controlled transparency in single-mode surface-corrugated waveguides is based on two principal points. First, the transport through such systems, as through any 1D disordered structure, is completely described by the ratio $L/L_{loc}$ of the waveguide length $L$ to the quantity $L_{loc}$ known as the localization length [5,6]. This concept is called ***one-parameter scaling***. The localization length is determined by the parameters of a waveguide, and by statistical characteristics of its rough surface. For large localization length, $L/L_{loc} \ll 1$, the waveguide is practically transparent and its transmittance $T$ is almost equal to unity. Otherwise, when $L_{loc}/L \ll 1$, the transmittance is exponentially small because of strong wave localization. Second, the inverse localization length is proportional to the lengthwise Fourier transform $W(k_x)$ of the binary correlator of random surface profile, $L_{loc}^{-1} \propto W(2k_1)$ where $k_1$ is the longitudinal wave number of single propagating waveguide mode. Therefore, if the roughness power spectrum $W(2k_1)$ abruptly vanishes within some interval of wave number $k_1$, then the localization length $L_{loc}$ diverges and the waveguide is fully transparent. Consequently, with the use of the methods developed in the theory of surface profile generation [7], we can construct such random surfaces that result in a complete transparency of waveguides within a predefined part of the allowed wave-number region [4]. A sharp transition between localized and ballistic transport at a given point implies the roughness power spectrum $W(k_x)$ having a discontinuity at this point. This means that the corrugated profile $\xi(x)$ of waveguide surface should have particular long-range correlations along the waveguide (along the coordinate $x$). Note that surfaces with the properties that give rise to rectangular roughness spectrum are not exotic. They have been recently fabricated in experimental studies of an enhanced backscattering [8].

**2 Quasi-1D waveguides with correlated surface disorder** Now, let us discuss transport properties of ***many-mode plane*** (quasi-1D) waveguides with correlated surface disorder. Below we show that in this case the role of long-range correlations is much more sophisticated in comparison with that discussed above for single-mode waveguides. The reason is that unlike the single-mode situation, the concept of one-parameter scaling is no more true for the transport through surface-corrugated many-mode waveguides. There are two points that should be stressed in this respect. First, the correlations discussed above result in suppression of the interaction between different propagating modes. This non-trivial fact turns out to be crucial for the reduction of a system of mixed channels with quasi-1D transport, to the coset of independent waveguide modes with purely 1D transport. Second, the same correlations lead to a complete transparency of each independent channel, similar to what happens in strictly single-mode geometry.

To start with, we should note that in the spirit of the Landauer's concept [9], in our case of many-mode waveguides the total average transmittance $<T>$ can be expressed as a sum of partial transmittances $T_n$ for every *n*th propagating normal mode (conducting channels),

$$<T> = \sum_{n=1}^{N_d} T_n .\qquad(1)$$

Here $N_d = [kd/\pi]$ is the total number of propagating modes determined by the integer part [?] of the ratio $kd/\pi$, and $d$ is the average waveguide width. The total wave number $k$ is equal to $\omega/c$ for a classical wave of frequency $\omega$, and to the Fermi wave number for electrons.

From the general theory of surface scattering [10] it follows that transmission properties of any *n*th conducting channel are determined by two attenuation lengths, the length $L_n^{(f)}$ of forward scattering and the backscattering length $L_n^{(b)}$. These scattering lengths are expressed by

$$\frac{1}{L_n^{(f)}} = \sigma^2 \frac{(\pi n/d)^2}{k_n d} \sum_{n'=1}^{N_d} \frac{(\pi n'/d)^2}{k_{n'} d} W(k_n - k_{n'}) ,\qquad(2)$$

$$\frac{1}{L_n^{(b)}} = \sigma^2 \frac{(\pi n/d)^2}{k_n d} \sum_{n'=1}^{N_d} \frac{(\pi n'/d)^2}{k_{n'} d} W(k_n + k_{n'}) ,\qquad(3)$$

with $\sigma$ being the root-mean-square roughness height and $k_n = \sqrt{k^2 - (\pi n/d)^2}$ the lengthwise wave number of *n*th propagating mode. These results can be obtained by the diagrammatic Green's function approach [10] as well as by the technique developed in Ref. [11]. Note that in a single-mode waveguide with $n = N_d = 1$ the sum over $n'$ contains only one term with $n' = 1$. In this case Eq. (3) for the backscattering length $L_1^{(b)}$ gives the discussed above value of the localization length $L_{loc}$ that is four times larger than $L_1^{(b)}$.

One can see from Eqs. (2) and (3) that, in general, both attenuation lengths are contributed by scattering of a given *n*th propagating mode into all other modes. This is the case when, for example, surface profiles $\xi(x)$ are either of white-noise type with constant power spectrum $W(k_x)$, or the random function with widely used fast decreasing Gaussian correlator (or, the same, with a slow decrease of its Fourier transform). Besides, these expressions manifest rather strong dependence on the mode index ***n***. Specifically, the larger the mode number ***n*** is, the smaller are the corresponding attenuation lengths and as a consequence, the stronger is the scattering of this mode into the others. As was shown in Ref. [12], even in the absence of correlations in $\xi(x)$ a very interesting phenomenon of the coexistence of ballistic, diffusive, and localized transport arises, which seems to be generic for propagation through many-mode waveguides with disordered surfaces. Namely, while lowest modes can be in the ballistic regime, the intermediate and highest propagating modes exhibit the diffusive and localized behavior, respectively. As a result, we come to the concept of the ***hierarchy of mode attenuation lengths*** instead of the one-parameter scaling.

Now let us demonstrate that the situation fundamentally changes when the surface roughness has specific long-range correlations. As an example, we consider the random surface profile $\xi(x)$ with the simplest power spectrum in the form of a "window function",

$$W(k_x) = (\pi/k_c)\Theta(k_c - |k_x|). \tag{4}$$

Here $\Theta(z)$ stands for the unit-step function and the characteristic wave number $k_c > 0$ is the controlling parameter to be specified below.

It is evidently that in the case under consideration the number of modes into which a given *n*th mode is scattered, i.e. the actual number of the summands of Eqs. (2) and (3), is entirely determined by the width $k_c$ of the rectangular spectrum (4). Moreover, if the distance $|k_n - k_{n\pm1}|$ between neighboring wave numbers is larger than the controlling width $k_c$,

$$|k_n - k_{n\pm1}| > k_c, \tag{5}$$

then the transitions between all modes are forbidden. In this case the sum over $n'$ in Eq. (2) for the forward scattering length $L_n^{(f)}$ contains only one term with $n' = n$ which describes intra-mode scattering only. At the same time, each term in the sum (3) is equal to zero so that the backscattering length is infinite, $L_n^{(b)} = \infty$.

From the above consideration a remarkable effect can be revealed. Namely, all the propagating modes with index ***n*** for which the condition (5) holds, are fully independent of other waveguide modes, in spite of their interaction with a rough surface. In other words, they represent a coset of 1D non-interacting channels. As is well known from the standard theory of 1D localization (see, e.g., Refs. [5,6]), the transmission through any 1D disordered structure is determined by the backscattering length only and does not depend on the forward scattering. Since the former is infinite for every such channel, its partial transmittance is equal to unity, $T_n = 1$. As for other propagating modes with index ***n*** that is in contradiction with the condition (5), they remain to be mixed by the surface scattering because the roughness power spectrum (4) for them is non-zero, $W(k_n - k_{n'}) = \pi/k_c$. Since these ***mixed modes*** have finite attenuation lengths, for large enough waveguide length $L$ they do not contribute to the total transmittance (1) and the latter is equal to the total number of ***independent transparent modes***

The further analytical treatment is allowed for large number of conducting channels

$$N_d = [kd/\pi] \approx kd/\pi \gg 1. \tag{6}$$

In this case the condition (5) is equivalent to the requirement $|\partial k_n/\partial n| > k_c$, which can be written in the explicit form

$$n > N_m \equiv \left[(kd/\pi)\{1 + (k_c d/\pi)^{-2}\}^{-1/2}\right], \tag{7}$$

where, as above, the square brackets stand for the integer part of the inner expression. Thus, all propagating modes with $n > N_m$ are independent and fully transparent, otherwise, they are mixed and characterized by finite attenuation lengths. Therefore, the integer $N_m$ should be regarded as the total number of mixed non-transparent modes while the total number of independent transparent channels is given by the difference $N_t = N_d - N_m$.

One can see from Eq. (7) that the numbers $N_m$ and $N_t$ of mixed non-transparent and independent transparent modes are determined by two parameters: the mode parameter $\alpha = kd/\pi$ and the dimensionless ***correlation parameter*** (CP) $\alpha_c = k_c d/\pi$. In the case of "weak" correlations when $\alpha_c \gg 1$, the number of mixed modes $N_m \approx [\alpha(1 - \alpha_c^{-2}/2)]$ is of the order of $N_d = [\alpha]$. Consequently, in this case the number of independent transparent modes $N_t = N_d - N_m$ is small, or there are no such modes at all. If the CP $\alpha_c$ tends to infinity ($\alpha_c \to \infty$) the rough surface profile becomes white-noise-like and, naturally, $N_m \to N_d$. The most appropriate case is when surface roughness is strongly correlated so that the CP is small, $\alpha_c \ll 1$. Then the number of mixed non-transparent modes $N_m \approx [\alpha_c \alpha]$ is much less than the total number of propagating modes $N_d$ and the number of transparent modes $N_t$ is large. When $\alpha_c$ decreases and becomes anomalously small ($\alpha_c < \alpha^{-1} \ll 1$), the number $N_m$ vanishes and all modes turn out to be independent and fully transparent. In this case the correlated disorder results in a perfect transmission of waves, in spite of their scattering from a rough surface.

Finally, let us briefly discuss the expression for the transmittance of many-mode waveguides with correlated surface roughness of the above kind,

$$<T> = [\alpha] - [\alpha(1+\alpha_c^{-2})^{-1/2}]. \qquad (8)$$

It is clear that the transmittance (8) reveals a step-wise dependence on the value of the mode parameter $\alpha = kd/\pi$. The analogous effect is known to occur for the conductance of quasi-1D ballistic (non-disordered) structures (see, e.g., Ref. [13]). However, in our surface-disordered model the step-wise dependence consists not only of the usual (ballistic) *steps up* from the first term in Eq. (8) but also of the *steps down*. The latter belong to the second term and are formed by the correlated surface scattering. The steps up arise for integer values of $\alpha = kd/\pi$ only. In contrast to this, the positions of the steps down are determined by the correlation parameter $\alpha_c = k_c d/\pi$ and can occur for non-integer values of $kd/\pi$. Evidently, within some intervals the steps of such different types cancel each other. Therefore, the experimental picture of the discussed dependence is expected to be very interesting and sophisticated.

**3 Summary** We have studied the role of long-range correlations in surface profiles for the transport through quasi-1D waveguides. It was found that the correlations that result in a complete transmission of waves in the case of one-mode waveguides, lead to a quite unexpected phenomenon when the number of modes is large. Specifically, we show that these long-range correlations give rise to the appearance of non-interacting 1D channels, the number of which is controlled by the correlation parameter. The remarkable point is that these channels turn out to be completely transparent in some range of frequency of incoming waves. As a result, the total transmission of waveguides can be significantly enhanced in com comparison with uncorrelated surface profiles (or profiles with Gaussian correlations of finite length). Moreover, the number of independent transparent channels can be as large as the total number of modes. In this case, the waveguides are fully transparent. This effect is directly related to the fact that for many-mode surface scattering there are many characteristic lengths that determine the total transport. In other words, the famous single-parameter scaling known to be held for bulk scattering, in our case is not valid. As a result, surface scattering transport through different channels can be separated by a proper choice of long-range correlations along surface profiles. Our study may find practical applications, for example, when fabricating of waveguides and superlattices with selective transport in a given frequency range of incoming waves. Although our consideration is based on the first order approximation (2-3), the higher correction terms can not change drastically the whole picture for the mobility edge transition. For this reason it is naturally to expect that global properties of the transport are correctly described by our theory. Specifically, transitions from metallic to localized regimes discussed above for the correlated disorder, are expected to be sharp enough in order to observe them experimentally.

**Acknowledgements** This work was partially supported by the Consejo Nacional de Ciencia y Tecnologia (CONACYT, Mexico) under Grants No. 34668-E, 36047-E and by the Universidad Autonoma de Puebla (BUAP, Mexico) under Grant II-104G02.

### References

[1] P. Carpena, P. Bernaola-Galvan, P. Ch. Ivanov, and H. E. Stanley, Nature (London) **418**, 955 (2002).
[2] F. M. Izrailev and A. Krokhin, Phys. Rev. Lett. **82**, 4062 (1999); A. A. Krokhin and F. M. Izrailev, Ann. Phys. (Leipzig) **8**, 153 (1999); F. M. Izrailev, A. A. Krokhin, and S. E. Ulloa, Phys. Rev. B **63**, 041102(R) (2001).
[3] U. Kuhl, F. M. Izrailev, A. A. Krokhin, and H.-J. Stoeckmann, Appl. Phys. Lett. **77**, 633 (2000).
[4] F. M. Izrailev and N. M. Makarov, Optics Letters **26**, 1604 (2001).

[5] I. M. Lifshits, S. A. Gredeskul, and L. A. Pastur, Introduction to the Theory of Disordered Systems (Wiley, New York, 1988).
[6] N. M. Makarov and I. V. Yurkevich, Zh. Eksp. Teor. Fiz **96**, 1106 (1989) [Sov. Phys. JETP **69**, 628 (1989)]; V. Freilikher, N. M. Makarov, and I. V. Yurkevich, Phys. Rev. B **41**, 8033 (1990).
[7] S. O. Rice, in Selected Papers on Noise and Stochastic Processes, ed. by N. Wax (Dover, 1954) p. 180.
[8] C. S. West and K. A. ODonnell, J. Opt. Soc. Am. A **12**, 390 (1995).
[9] R. Landauer, Physica Scripta **T42**, 110 (1992).
[10] F. G. Bass, I. M. Fuks, Wave Scattering from Statistically Rough Surfaces (Pergamon, New York, 1979).
[11] A. R. McGurn and A. A. Maradudin, Phys. Rev. B **30**, 3136 (1984).
[12] J. A. Sánchez-Gil, V. Freilikher, I. V. Yurkevich, and A. A. Maradudin, Phys. Rev. Lett. **80**, 948 (1998); J. A. Sánchez-Gil, V. Freilikher, A. A. Maradudin, and I. V. Yurkevich, Phys. Rev. B **59**, 5915 (1999).
[13] B. J. van Wees, H. van Houten, C. W. J. Beenakker, et al, Phys. Rev. Lett. **60**, 848 (1988).